\documentclass[aps,pra,amsmath,amssymb,amsthm,floatfix,nofootinbib,notitlepage,superscriptaddress,10pt]{revtex4-1}
\usepackage{esvect,url}
\pdfoutput=1
\newcommand{\lin}{L}
\usepackage{amsthm}
\usepackage{latexsym}
\usepackage{amsfonts}
\usepackage{bbm,dsfont}
\usepackage{graphicx,xcolor}
\usepackage{braket}
\usepackage{hyperref}
\usepackage[normalem]{ulem}
\usepackage{braket}
\definecolor{bottle_green}{RGB}{0,106,78}
\definecolor{celadon_green}{RGB}{47,132,124}
\definecolor{emerald}{RGB}{80,220,100}
\definecolor{jade}{RGB}{0,168,107}
\hypersetup{colorlinks,
linkcolor=bottle_green,
citecolor=celadon_green,
urlcolor=emerald,
anchorcolor=jade}
\newcommand{\Tr}[2]{\mathrm{Tr}_{#1}\left[ {#2} \right]}

\newcommand{\cqstate}{\varrho}

\begin{document}
%%%%%%%%%%%%%%
\title{The two classes of hybrid classical-quantum dynamics}
%A classical-quantum Pawula theorem }
\author{Jonathan Oppenheim}
\affiliation{Department of Physics and Astronomy, University College London, Gower Street, London WC1E 6BT, United Kingdom}
\author{Carlo Sparaciari}
\affiliation{Department of Physics and Astronomy, University College London, Gower Street, London WC1E 6BT, United Kingdom}
\author{Barbara \v{S}oda}
\affiliation{Dept. of Physics,
University of Waterloo, Waterloo, Ontario, Canada}
\affiliation{Perimeter Institute for Theoretical Physics, Waterloo, Ontario, Canada}
\affiliation{Department of Physics and Astronomy, University College London, Gower Street, London WC1E 6BT, United Kingdom}
\author{Zachary Weller-Davies}
\affiliation{Perimeter Institute for Theoretical Physics, Waterloo, Ontario, Canada}
\affiliation{Department of Physics and Astronomy, University College London, Gower Street, London WC1E 6BT, United Kingdom}

%%%%%%%%%%%%%%
\date{\today}
%%%%%%%%%%%%%%

%%%%%%%%%%%%%%
%%%%%%%%%%%%%%
%%%%%%%%%%%%%%

\begin{abstract}
Coupling between quantum and classical systems is consistent, provided the evolution is linear in the state space, preserves the split of systems into quantum and classical degrees of freedom, and preserves probabilities. The evolution law must be a completely positive and norm preserving map. We prove that if the dynamics is memoryless, there are two classes of these dynamics, one which features finite sized jumps in the classical phase space and one which is continuous. We find the most general form of each class of classical-quantum master equation. 
This is achieved by applying the complete positivity conditions using a generalized Cauchy-Schwartz inequality applicable to classical-quantum systems. The key technical result is a generalisation of the Pawula theorem.
\end{abstract}
%%%%%%%%%%%%%%
\maketitle
%%%%%%%%%%%%%%
%%%%%%%%%%%%%%
\section{Introduction}
Examples of consistent classical-quantum (CQ) dynamics have been known since the mid 90's \cite{blanchard1995event,diosi1995quantum}. These are of interest when we want to study the back-reaction of a quantum system on another which can be treated classical. In addition to this effective theory point of view \cite{alicki2003completely}, one could also study them as an alternative to quantum gravity \cite{diosi2011gravity,poulinKITP2,oppenheim_post-quantum_2018, Oppenheim:2020ogy}. Dynamics in the case where a classical degree of freedom such as the Newtonian potential directly encode measurements of the quantum system have also been studied via continuous measurement and feedback  \cite{Kafri_2014, Kafri_2015,tilloy2016sourcing, tilloy2017principle,HalliwellDiosi}, corresponding to Lindbladian evolution on the quantum system.  
CQ dynamics must be
%allows for independent classical degrees of freedom, and is
 linear in the state, completely positive, preserves the classical quantum split and preserves normalization of the state, ensuring that measurement probabilities remain positive and sum to 1. If the classical system is represented as a finite dimensional Hilbert space, then the most general form of the dynamics was shown \cite{PoulinPC2} to have the form of \cite{blanchard1995event}. When the phase space is continuous, one can use a CQ generalization of Krauss theorem \cite{kraus} to find the most general form of this dynamics for bounded Lindblad operators \cite{oppenheim_post-quantum_2018}.

In general, many classical-quantum dynamics appear to involve finite sized jumps in the classical phase space \cite{oppenheim_post-quantum_2018, Oppenheim:2020wdj}, yet CQ dynamics which appear continuous have also been found \cite{diosi1995quantum, diosi2011gravity}. However, the most general form of this dynamics is not known \cite{diosi2014hybrid, galley2021nogo}. In this note we shall remedy this,  fully characterizing continuous classical-quantum master equations. We find the most general continuous CQ dynamics takes the form 
\begin{align}\nonumber
\frac{\partial \cqstate(z,t)}{\partial t} & =  \sum_{n=1}^{n=2}(-1)^n\left(\frac{\partial^n }{\partial z_{i_1} \dots \partial z_{i_n} }\right) \left( D^{00}_{n, i_1 \dots i_n} \cqstate(z,t) \right) + \frac{\partial }{\partial z_{i}} \left( D^{0\alpha}_{1, i} \cqstate(z,t) L_{\alpha}^{\dag} \right)  + \frac{\partial }{\partial z_{i}} \left( D^{\alpha 0 }_{1, i} L_{\alpha} \cqstate(z,t) \right)   \\ \label{eq: continuous}
& -i[H(z), \cqstate(z,t)] + D_0^{\alpha \beta}(z) L_{\alpha} \cqstate(z) L_{\beta}^{\dag} - \frac{1}{2} D_0^{\alpha \beta} \{ L_{\beta}^{\dag} L_{\alpha}, \cqstate(z) \}_+,
\end{align}
where $2 D_{2}^{00} \succeq D_1 D_0^{-1} D_1^{\dag}$ and $(\mathbb{I}- D_0 D_0^{-1})D_1 =0$. Here, and throughout, $D^{-1}_0 $ is the generalized inverse of the positive semi-definite Lindbladian coupling $D_0^{\alpha \beta}$, $D_1$ is a matrix in both $\alpha, i$ indices with entries $D_{1,i}^{0 \alpha}$, which encodes the strength of the CQ back-reaction, and $D_2^{00}$ is a matrix in $i,j$ with entries $D_{2,ij}^{00}$, which represents the necessity of diffusion in the classical phase space. We show that CQ dynamics which is not of this form has finite sized jumps in phase space. The master equation of \cite{diosi1995quantum} is an example of this form, and for completeness we give a simple example of coupled classical and quantum harmonic oscillators in Appendix \ref{app: example}.

The measurement and feedback models of \cite{Kafri_2014, Kafri_2015,tilloy2016sourcing, tilloy2017principle,HalliwellDiosi} are not of this form, since the Newtonian potential is directly sourced by a weak measurement outcome, leading to the classical degrees of freedom evolving discontinuously and the quantum degrees of freedom evolving via a Lindblad equation. Other continous measurement and feedback models can be put into this form \cite{UCLContinousUnraveling}.

 In classical dynamics, we can write the master equation in terms of the moments of the transition probability amplitude via the Kramers-Moyal expansion~\cite{kramers1940brownian,moyal1949stochastic,risken1989fpe}. Positivity of the dynamics means the transition amplitude must be positive, which can then be used to derive constraints on the allowed moments in the moment expansion. Of particular relevance is the Pawula theorem \cite{pawula1967rf}, which states that the moment expansion either stops after the first or second moments, or else it must contain an infinite number of terms; in the former case, this restricts continuous dynamics to the well-known Fokker-Planck equation \cite{risken1989fpe}.\footnote{It is important to note that if one truncates the series after $n$ terms with $n \geq 3$, the resulting equation, although not positive, can still be used as an approximation to the dynamics in an appropriate regime. Indeed, one might attain a better approximation of certain classical dynamics by using an approximation that is not positive; one just has to be careful about the validity of the approximation \cite{risken1989fpe}.}

Our main technical result is a proof of a classical-quantum version of the Pawula theorem, which follows from a combined CQ Cauchy-Schwarz inequality, Equation \eqref{eq: bigineq}. We find that in order for a non-trivial classical-quantum interaction to be completely positive, the classical-quantum moment expansion must either contain an infinite number of terms, or it must be of the form Equation \eqref{eq: continuous}. Infinite moments are indicative of a jump process and so we prove that Equation \eqref{eq: continuous} is the unique, CQ master equation which has continuous trajectories in phase-space. In \cite{UCLContinousUnraveling} we show explicitly how one can unravel the continuous master equation in terms of coupled stochastic differential equations, and give this an interpretation in terms of the classical system continuously measuring the quantum system.

A natural consequence of the CQ Pawula theorem, is that in order for classical-quantum  dynamics to be completely positive one must have a term representing pure Lindbladian evolution on the quantum state. In other words, the nature of completely positive dynamics necessarily results in the classical degrees of freedom inducing decoherence on the quantum state. Classicality induces classicality.

\section{Classical-Quantum dynamics} \label{sec: cqdnamics}
Let us first briefly review the general map and master equation governing classical-quantum dynamics. The classical degrees of freedom are described by a configuration space $\mathcal{M}$ and we shall generically denote elements of the classical space by $z$. For example, we could take the classical degrees of freedom to be position and momenta in which case $\mathcal{M}= \mathbb{R}^{2}$ and $z= (q,p)$. The quantum degrees of freedom are described by a Hilbert space $\mathcal{H}$.  Given the Hilbert space, we denote the set of positive semi-definite operators with trace at most unity as $S_{\leq 1}(\mathcal{H})$. Then the CQ object defining the state of the CQ system at a given time is a map $\cqstate : \mathcal{M} \to  S_{\leq 1}(\mathcal{H})$ subject to a normalization constraint $\int_{\mathcal{M}} dz \Tr{\mathcal{H}}{\cqstate} =1$. To put it differently, we associate to each classical degree of freedom a sub-normalized density operator, $\cqstate(z)$, such that $\Tr{\mathcal{H}}{\cqstate} = p(z) \geq 0$ is a normalized probability distribution over the classical degrees of freedom and $\int_{\mathcal{M}} dz \cqstate(z) $ is a normalized density operator on $\mathcal{H}$. 

In the case where the classical degrees of freedom are taken to be discrete, it has been shown~\cite{PoulinPC2} that any bounded dynamics mapping trace-class operators and CQ states onto themselves, if taken to be linear and Markovian, will be completely positive if and only if it can be written in the form
 \begin{equation}\label{eq: CPmap}
 \cqstate(z,t+ \delta t) =  \int dz' \Lambda(z|z',\delta t) ( \cqstate(z',t)) = \int dz'\sum_\mu \Lambda^{\mu}(z|z',\delta t) \lin_{\mu}(z,z',\delta t) \cqstate(z',t) \lin_{\mu}^{\dag}(z,z',\delta t) ,
\end{equation}  
where $ \Lambda(z|z',\delta t)$ is a completely positive map for each $z,z'$, the $L_{\mu}(z,z',\delta t)$ are an orthogonal basis of operators and $\Lambda^{\mu}(z|z',\delta t)$ is positive for each $z,z'$. 
The normalization of probabilities requires
\begin{equation}\label{eq: prob}
\int dz  \sum_\mu\Lambda^{\mu}(z|z',\delta t) \lin_{\mu}^{\dag}(z,z',\delta t)\lin_{\mu}(z,z',\delta t) =\mathbb{I}.
\end{equation}
The choice of basis $L_\mu$ is arbitrary, although there may be one which allows for unique trajectories \cite{Oppenheim:2020wdj}.
Equation \eqref{eq: CPmap} can be viewed as a generalisation of the Kraus decomposition theorem. However, when the classical degrees of freedom are taken to live in a continuous configuration space, we need to be a little more careful, since $\cqstate(z)$ may only be defined in a distributional sense; for example, $\cqstate(z) = \delta(z, \bar{z}) \cqstate( \bar{z})$. In this case \eqref{eq: CPmap} is completely positive if  $\int dz dz' P_{\mu}(z,z') \Lambda^{\mu}(z|z') \geq 0$ for any positive $P_{\mu}(z,z')$. We show in appendix \ref{sec: proof} that CQ dynamics with continuous classical degrees of freedom will be positive only if it can be written in the form of \eqref{eq: CPmap}. 

One can derive the general form of CQ master equation by performing a short time expansion of \eqref{eq: CPmap} in the case when the $L_\mu$ are bounded \cite{oppenheim_post-quantum_2018}. To do so, we first introduce an arbitrary basis of traceless Lindblad operators on the Hilbert space, $L_{\mu} = \{I, L_{\alpha}\}$, defined in terms of the operators in \eqref{eq: CPmap} via $L_{\mu} = U_{\mu \nu}(z,z',\delta t) L_{\nu}(z,z',\delta t)$. This enables us to write
\begin{equation}\label{eq: CPmap4}
 \cqstate(z,t+ \delta t) = \sum_{\mu\nu} \int dz' \Lambda^{\mu \nu}(z|z',\delta t) \lin_{\mu} \cqstate(z',t) \lin_{\nu}^{\dag},
\end{equation}  
where we define  $\Lambda^{\mu \nu}(z|z',\delta t) = U^{\dag}_{\mu \sigma}  \Lambda^{\sigma} U_{\sigma \nu}$, which is a positive matrix in $\mu \nu$. Henceforth, we will adopt the Einstein summation convention so that we can drop $\sum_{\mu\nu}$ with the understanding that equal upper and lower indices are presumed to be summed over.

At $\delta t=0$ we know \eqref{eq: CPmap4} is the identity map, which tells us that $\Lambda^{00} (z|z', \delta t=0) = \delta(z,z')$ and $L_0(z,z', \delta t =0) = \mathbb{I} $. Looking at the short time expansion coefficients, by Taylor expanding in $\delta t \ll 1$, we can write
\begin{align}\label{eq: shortime}
&\Lambda^{\mu \nu}(z|z',\delta t) = \delta^{\mu}_{0} \delta^{ \nu}_{0} \delta(z,z') + W^{\mu \nu}(z|z') \delta t + O(\delta t^2).
\end{align}
By substituting the short time expansion coefficients into \eqref{eq: CPmap4} and taking the limit $\delta t \to 0$ we can write the master equation in the form 
\begin{align}
\frac{\partial \cqstate(z,t)}{\partial t} =  \int dz' \ W^{\mu \nu}(z|z') L_{\mu} \cqstate(z') L_{\nu}^{\dag} - \frac{1}{2}W^{\mu \nu}(z) \{ L_{\nu}^{\dag} L_{\mu}, \cqstate \}_+,
\label{eq: cqdyngen}
\end{align}
where $\{, \}_+$ is the anti-commutator, and preservation of normalisation under the trace and $\int dz$ defines
\begin{equation}
W^{\mu  \nu}(z) = \int \mathrm{d} \Delta W^{\mu \nu}(z+ \Delta| z).
\end{equation}
We see the CQ master equation is a natural generalisation of the Lindblad equation and classical rate equation  in the case of classical-quantum coupling. We give a more precise interpretation of the different terms arising in section \ref{sec: Kramers}, where we review the Kramers-Moyal expansion of the master equation. The positivity conditions from \eqref{eq: CPmap} transfer to positivity conditions on the master equation via \eqref{eq: shortime}, in the case where the dynamics is Markovian. In the non-Markovian case, which we do not consider here, the $W^{\mu  \nu}(z)$ are less constrained \cite{Hall2014, Breuer2016}. We can write the positivity conditions in an illuminating form by writing the short time expansion of the transition amplitude $\Lambda^{\mu \nu}(z|z',\delta t)$, as defined by equation \eqref{eq: shortime}, in block form
\begin{equation} \label{eq: block1}
\Lambda^{\mu \nu}(z|z',\delta t) 
= \begin{bmatrix}
    \delta(z,z') + \delta t W^{00}(z|z')      & \delta t W^{0\beta}(z|z') \\
    \delta t W^{\alpha 0}(z|z')    & \delta t W^{\alpha \beta} (z|z')  \\
\end{bmatrix} + O(\delta t^2).
\end{equation}
The dynamics will be positive if and only if $\Lambda^{\mu \nu}(z|z',\delta t)$ is a positive matrix. From this we immediately deduce that $\Lambda^{00}(z|z') = \delta(z,z') + \delta t W^{00}(z|z')$ must be positive, as well as the matrix $\Lambda^{\alpha \beta}(z|z',\delta t) = \delta t W^{\alpha \beta} (z|z')$. Furthermore, if either of $W^{\alpha \beta}(z|z')$ or $W^{00}(z|z')$ vanish, then so must $W^{0 \alpha}(z|z')$, except for its $z=z'$ component which generates pure Hamiltonian evolution. This tell us in order to have non-trivial CQ coupling we must have a non-zero $W^{\alpha \beta}(z|z')$. 

As we shall see, this has important consequences for CQ dynamics. It is useful to note that when the classical degrees of freedom are discrete, the Schur complement -- assuming $W^{00}(z|z')$ is non-vanishing -- informs us the matrix $\Lambda^{\mu \nu}(z|z',\delta t)$ will be positive if and only if $W^{00}(z|z') W^{\alpha \beta}(z|z') - W^{0 \beta}(z|z') W^{ \alpha 0}(z|z') \succeq 0$ is a positive matrix in $\alpha \beta$ for all $z \neq z'$. In the continuous case we have to be a little more careful, since the components $\Lambda^{\mu \nu}(z|z')$ may only be defined in a distributional sense, and exploring the positivity conditions in this case is one of the main goals of the present paper.

It is possible to introduce an arbitrary basis of Lindblad operators $\bar{L}_{\mu}$ and appropriately redefine the couplings $W^{\mu \nu}(z|z')$ in \eqref{eq: cqdyngen}. For most purposes, we shall work with a basis of traceless Lindblad operators $(I, L_{\alpha})$; this is sufficient since any CQ master equation is completely positive if and only if it can be brought to the form in \eqref{eq: cqdyngen}, where the matrix \eqref{eq: block1} is positive. The exception is in section \ref{sec: CQPawulasec} where we use a specific choice of basis such that \eqref{eq: block1} does not contain the off-diagonal terms $W^{0 \alpha}(z|z')$, except for its $z = z'$ component.

We shall often deal with superoperators and it shall prove useful to occasionally double the quantum degrees of freedom using the vectorization  map~\cite{nielsen00}. We do so by representing the CQ density operators $\cqstate(z)$ as vectors by stacking the columns, i.e, sending $ |i \rangle \langle j | \to | j \rangle \otimes | i \rangle$.  We denote the vectorized form as $\vec{\cqstate}(z) $. Then, superoperators are matrices acting on the stacked vector $\vec{\cqstate}(z)$, for example
\begin{equation}
\vec{\cqstate}(z,t+ \delta t) = \int dz' \Lambda^{\mu \nu}(z|z',\delta t) (\bar{\lin}_{\nu} \otimes \lin_{\mu}) \vec{\cqstate}(z',t) = \int dz' \Lambda^{vec}(z|z',\delta t)( \vec{\cqstate}(z')),
\end{equation}
where we write $vec$ to remind us that we should view the superoperator as a matrix on the doubled Hilbert space. This is particularly useful since it allows us to identify the components of the superoperator in any orthogonal basis of operators $(\bar{\lin}_{\nu} \otimes \lin_{\mu})$ via
\begin{equation}\label{eq: basis}
\Lambda^{\mu \nu}(z|z',\delta t) = \Tr{}{ ( \bar{L}_{\nu} \otimes L_{\mu})^{\dag} \Lambda^{vec}(z|z',\delta t)}.
\end{equation}

\subsection{Master equation and short time moment expansion coefficients}\label{sec: Kramers}
In order to study the positivity conditions it is first useful to perform a moment expansion of the dynamics in a classical-quantum version of the Kramers-Moyal expansion~\cite{oppenheim_post-quantum_2018}. In classical Markovian dynamics, the Kramers-Moyal expansion relates the master equation to the moments of the probability transition amplitude and proves to be useful for a multitude of reasons. Firstly, the moments are related to observable quantities; for example, the first and second moments of the probability transition amplitude characterize the amount of drift and diffusion in the system. Secondly, the positivity conditions on the master equation transfer naturally to positivity conditions on the moments, which we can then relate to observable quantities. In the classical-quantum case, we shall perform a short time moment expansion of the transition amplitude $\Lambda^{\mu \nu}(z|z',\delta t) $ and then show that the master equation can be written in terms of these moments. We then relate the moments to observational quantities, such as the decoherence of the quantum system and the diffusion in the classical system. 

We work with the form of the dynamics in \eqref{eq: cqdyngen}, using an arbitrary orthogonal basis of Lindblad operators $L_{\mu} = \{I, L_{\alpha}\}$. We take the classical degrees of freedom $\mathcal{M}$ to be $d$ dimensional, $z= (z_1, \dots z_d)$, and we label the components as $z_i$, $i \in \{1, \dots d\}$. We begin by introducing the moments of the transition amplitude 
\begin{equation}\label{eq: moments}
M^{\mu \nu}_{n, i_1 \dots i_n} (z',\delta t) = \int dz \ \Lambda^{\mu \nu}(z|z',\delta t) (z-z')_{i_1} \dots (z-z')_{i_n},
\end{equation}
where $\Lambda^{\mu \nu}(z|z',\delta t)$ are the components of the dynamics of the CP map in the basis $L_{\mu} = \{I, L_{\alpha}\}$, as defined in \eqref{eq: basis}. The subscripts $i_j \in  \{1, \dots d\}$ label the different components of the vectors $(z-z')$. For example, in the case where $d=2$ and the classical degrees of freedom are position and momenta of a particle, $z=(z_1,z_2) = (q,p)$, then we have $(z-z') = (z_1- z'_1, z_2 - z_2') = ( q-q',p-p')$. The components are then given by $(z-z')_1 = (q-q')$ and $(z-z')_2 = (p-p')$.  $M^{\mu \nu}_{n, i_1 \dots i_n} (z',\delta t)$ is seen to be an $n$'th rank tensor with $d^n$ components.

We define the characteristic function, which is the Fourier transform of the transition amplitude
\begin{equation}
C^{\mu \nu}(u,z',\delta t) = \int dz e^{iu \cdot (z-z')} \Lambda^{\mu \nu}(z,z') = \sum_{n=0}^{\infty} \frac{(i^n) u_{i_1} \dots u_{i_n}}{n!} M^{\mu \nu}_{n, i_1 \dots i_n}(z',\delta t) .
\end{equation}
Taking the inverse Fourier transform, we can relate the transition amplitude to its moments
\begin{equation}
\Lambda^{\mu \nu}(z|z',\delta t) =\int du \ e^{-i u (z-z')}  C^{\mu \nu}(u,z',\delta t)  = \sum_{n=0}^{\infty}  \frac{M^{\mu \nu}_{n, i_1 \dots i_n}(z',\delta t)}{n!}   \frac{1}{(2 \pi)^d} \int du \ \ e^{-i u (z-z')} (i^n) u_{i_1} \dots u_{i_n},
\end{equation} 
which, using the definition of the delta distribution, we can write as
\begin{equation}\label{eq: transamp}
 \Lambda^{\mu \nu}(z|z',\delta t) = \sum_{n=0}^{\infty}\frac{1}{n!} M^{\mu \nu}_{n, i_1 \dots i_n}(z',\delta t)  \left(\frac{\partial^n }{\partial z_{i_1}' \dots \partial z_{i_n}' }\right) \delta(z,z') .
\end{equation}

Looking at the short time expansion coefficients of $\Lambda^{\mu \nu}(z|z',\delta t)$, as defined in \eqref{eq: shortime}, we have
\begin{equation}
M^{\mu \nu}(z',\delta t)_{n, i_1 \dots i_n} = \delta^{\mu}_0 \delta^{\nu}_0 + \delta t \int dz W^{\mu \nu}(z|z')(z-z')_{i_1} \dots (z-z')_{i_n} \equiv  \delta^{\mu}_0 \delta^{\nu}_0 + \delta t n! D^{\mu \nu}_{n, i_1 \dots i_n}(z')  + O(\delta t^2),
\label{eq: defM}
\end{equation} 
where we define quantity $D^{\mu \nu}(z')_{n, i_1 \dots i_n}$ via
\begin{equation}
   D^{\mu \nu}_{n, i_1 \dots i_n}(z'):= \frac{1}{n!}\int dz W^{\mu \nu}(z|z')(z-z')_{i_1} \dots (z-z')_{i_n} .
\end{equation}
 We shall occasionally find it useful to refer to the moments as $D_n(z')$, by which we mean the object with components $D^{\mu \nu}(z')_{n, i_1 \dots i_n}$. Substituting the short time moment coefficients back into \eqref{eq: transamp}, taking the limit $\delta t \to 0$ and using the probability preserving condition in \eqref{eq: prob}, we can write the master equation in the form
\begin{align}\label{eq: expansion}\nonumber
\frac{\partial \cqstate(z,t)}{\partial t} &= \sum_{n=1}^{\infty}(-1)^n \left(\frac{\partial^n }{\partial z_{i_1} \dots \partial z_{i_n} }\right) \left( D^{00}_{n, i_1 \dots i_n}(z,\delta t) \cqstate(z,t) \right)\\ \nonumber
&  -i  [H(z),\cqstate(z)  ] + D_0^{\alpha \beta}(z) L_{\alpha} \cqstate(z) L_{\beta}^{\dag} - \frac{1}{2} D_0^{\alpha \beta} \{ L_{\beta}^{\dag} L_{\alpha}, \cqstate(z) \}_+ \\ 
& + \sum_{\mu \nu \neq 00} \sum_{n=1}^{\infty}(-1)^n  \left(\frac{\partial^n }{\partial z_{i_1} \dots \partial z_{i_n} }\right)\left( D^{\mu  \nu}_{n, i_1 \dots i_n}(z) \lin_{\mu} \cqstate(z,t) \lin_{\nu}^{\dag} \right),
\end{align}
where we have defined the Hermitian operator $H(z)= \frac{i}{2}(D^{\mu 0}_0 L_{\mu} - D^{0 \mu}_0 L_{\mu}^{\dag}) $ (which  is Hermitian since $D^{\mu 0}_0 = D^{0 \mu *}_0$). We see the first line of \eqref{eq: expansion} describes purely classical dynamics, and is fully characterised by the moments of the identity component of the dynamics $\Lambda^{00}(z|z')$. The second line describes pure quantum Lindbladian evolution described by
the zeroth moments of the components $\Lambda^{\alpha 0}(z|z'), \Lambda^{\alpha \beta}(z|z')$; specifically the (block) off diagonals, $D^{\alpha 0}_0(z)$, describe the pure Hamiltonian evolution, whilst the
components $D^{\alpha \beta}_0(z)$ describe the dissipative part of the pure quantum evolution. Note that the Hamiltonian and Lindblad couplings can depend on the classical degrees of freedom so the second line describes action of the classical system on the quantum one. The third line contains the non-trivial classical-quantum back-reaction, where changes in the distribution over phase space are induced and accompanied by changes in the quantum state.

In classical Markovian dynamics, the moments of the short time expansion of the probability transition amplitude $P(z|z',\delta t)$ are useful since they are usually related to observable quantities. For example, the first moment characterizes the drift in the system, whilst the second moment typically characterizes diffusion. In the CQ case, we have similar interpretations. For example, it is seen from Equation \eqref{eq: expansion} that the zeroth moments characterize the pure quantum evolution. In particular, $D^{\alpha \beta}_0(z)$ characterizes the amount of decoherence on the quantum system. As we shall see, in order to have a non-trivial classical-quantum dynamics, positivity demands $D^{\alpha \beta}_0(z) \neq 0$ and so the classical system forces decoherence upon the quantum system. To give interpretations to the higher order moments, consider starting in a state of certainty in phase space, $\cqstate(z,t) =\delta(z,\bar{z}) \sigma$, where $\sigma$ is a normalized quantum state, and after some short time $\delta t$ measuring the classical observable $(z- \bar{z})^n$, $ n \geq 1$. In this case we find 
\begin{equation}
\int dz (z- \bar{z})^n \Tr{}{\cqstate(z,t+ \delta t)} = \delta t n ! ( D^{\mu \nu}_n(\bar{z}) \Tr{}{ L_{\nu}^{\dag} L_{\mu} \sigma }).
\label{eq:z_to_the_n}
\end{equation}
Hence we see the coefficients $ D^{\mu \nu}_n(z)$ (for $\mu \nu \neq 00$) characterize the back-reaction of the quantum system on the classical system in the presence of non-trivial CQ coupling. As we shall now prove, in order to have non-trivial classical-quantum coupling there must be infinitely many terms $D_n^{\alpha \beta}$, or else the dynamics must be of the form in Equation \eqref{eq: continuous}, which is the unique continuous CQ master equation.
\section{A classical-quantum Cauchy-Schwarz inequality}\label{sec: pos}
We know that the dynamics in \eqref{eq: expansion} will be positive so long as the transition amplitude $\Lambda^{\mu \nu}(z|z', \delta t) $ is a positive matrix and that positivity of $\Lambda^{\mu \nu}(z|z', \delta t) $ transfers naturally to positivity conditions on the short time moment expansions defined in equation \eqref{eq: shortime}; for example, by considering the block form of \eqref{eq: block1}. We first note that the pure classical positivity condition, given by the $00$ component $\Lambda^{00}(z|z',\delta t) = \delta(z,z') + \delta t W^{00}(z|z')$, leads to the well known Pawula theorem of classical Markovian dynamics; if any even moment $D^{00}_n$ vanishes then all moments with $n \geq 3$ must also vanish. For the unfamiliar reader, we include this proof in Appendix \ref{sec: pawulaproof}.

For the classical-quantum interaction to be completely positive, Equation \eqref{eq: block1} tells us $W^{\alpha \beta}(z|z')$ must be a positive matrix in $\alpha \beta$. We shall now use this fact to derive a family of inequalities which the moments must satisfy, which will in turn enable us to prove a strengthened version of the Pawula theorem to CQ dynamics. In particular, we show that there are two classes of CQ master equations: we either have infinitely many terms in the moment expansion or else the dynamics is given by that of Equation \eqref{eq: continuous}. Having infinitely many terms in the Kramers-Moyal expansion is characteristic of a jump process, therefore the only phase space continuous CQ master equation is given by \eqref{eq: continuous}. We further show that we \textit{must} have a non-zero pure decoherence term; the completely positivity of the CQ interaction necessarily causes the classical system to induce decoherence on the quantum system. 

We now derive a Cauchy-Schwartz like inequality, Equation \eqref{eq: bigineq}, applicable to any CQ map which is completely positive, and we use it to derive a set of inequalities relating the moments \ in \eqref{eq: expansion}. We first note that since $W^{\alpha \beta}(z|z')$ is a positive matrix, $W(z|z')( \cqstate(z')) = W^{\alpha \beta}(z|z') L_{\alpha} \cqstate(z') L_{\beta}^{\dag}$ defines a completely positive operator.  It will prove useful to use the vectorization map \eqref{eq: basis} to write the expansion coefficients $D^{\alpha \beta}_n(z)$ appearing in the dynamics of \eqref{eq: expansion} in terms of the components of the completely positive operator $W(z|z')$. Explicitly, 
\begin{align}
D^{\alpha \beta}_{n,i_1 \dots i_n}(z') L_{\alpha} \cqstate(z') L_{\beta}^{\dag} &= \frac{1}{n!}\int dz \ W^{\alpha \beta}(z|z') L_{\alpha} \cqstate(z') L_{\beta}^{\dag} (z-z')_{i_1} \dots (z-z')_{i_n}
\label{eq:genmoments}
\\
&=  \frac{1}{n!}\int dz \Tr{}{ (\bar{L}_{\beta} \otimes L_{\alpha} ) ^{\dag} W^{vec}(z|z' )}  L_{\alpha} \cqstate(z') L_{\beta}^{\dag}  (z-z')_{i_1} \dots (z-z')_{i_n}.
\end{align}
We could equally well write the completely positive operator $W(z|z')$ in terms of a different basis and it will prove useful to do so. To that end, given an arbitrary basis on the underlying Hilbert space $\{ |a \rangle\}$, we define the natural basis of operators on the Hilbert space $E_{ab}$, via $E_{ab} = |a \rangle \langle b |$. In this basis
\begin{equation}
D^{\alpha \beta}_{n, i_1 \dots i_n} (z) L_{\alpha} \cqstate L_{\beta}^{ \dag}  = D^{ab cd}_{n,  i_1 \dots i_n}(z) E_{ca} \cqstate E_{bd},
\end{equation}
where as in Equation \eqref{eq:genmoments}
\begin{align}
D^{abcd}_{ n,  i_1 \dots i_n}(z') := \int dz \frac{1}{  n!} \Tr{}{ (E_{db} \otimes E_{ca} )^{\dag} W^{vec}(z|z')} (z-z')_{i_1} \dots (z-z')_{i_n} \,\,\, .
\label{eq:momentsD}
\end{align}
Now, let us prove a generalised form of the Cauchy-Schwartz inequality that we can use for the case of hybrid classical-quantum theories. It will take the form
\begin{equation}
    \begin{split}
\label{eq: bigineq}
 \int d \Delta \Tr{\mathcal{H}}{ f(\Delta)^{\dag} f(\Delta) T(\Delta)}&  \int d \Delta \Tr{\mathcal{H}}{ g(\Delta)^{\dag} g(\Delta)  T(\Delta) } \\ 
& \geq  \int d\Delta \Tr{\mathcal{H}}{ f^{\dag}(\Delta) g(\Delta) T(\Delta)} \int d\Delta \Tr{\mathcal{H}}{ g^{\dag}(\Delta) f(\Delta) T(\Delta)},
\end{split}
\end{equation} 
and it holds for any completely positive operator $T(\Delta)$ and arbitrary CQ operators $g(\Delta), f(\Delta)$. The above relation is easily derived by rearranging
\begin{equation}
\int d\Delta d\Delta'    \Tr{A,B}{(f_A(\Delta) g_B(\Delta') - g_A(\Delta) f_B(\Delta'))^{\dag} (f_A(\Delta) g_B(\Delta') - g_A(\Delta) f_B(\Delta')) T_A(\Delta)T_B(\Delta')  } 
\label{eq: ineqder}
\end{equation}
which is certainly positive, owing to the fact that each map $T_{A/B}(\Delta)$ acting on it's share of a positive operator, is a completely positive map. Using \eqref{eq: bigineq} with  
\begin{equation} \label{eq: Pawhybrid}
\begin{split}
 T(z+ \Delta,z) & = W^{vec}(z+ \Delta,z) \\
  f(\Delta) = ( E_{bb} \otimes E_{aa}) \Delta_{i_1} \dots \Delta_{i_n} &, \ \ g(\Delta)= ( E_{bd}  \otimes E_{ac} )\Delta_{i_{n+m}} \dots \Delta_{i_{2n+2m} },
\end{split}
\end{equation}
and then integrating over $z$, we find the inequalities on the moments arising in the CQ equation
\begin{equation}\label{eq: momineq}
(2n!)(2n+2m)! D^{ab ab}_{ 2n, i_1 i_1 \dots i_n i_n} D^{cd cd}_{2n+2m, i_{n+m}i_{n+m} \dots i_{2n+2m}i_{2n+2m}} \geq |(2n+m)!D^{ab cd}_{ 2n+m, i_1 \dots i_{2n+m}} |^2,
\end{equation}
where we have used $D^{ab cd} = (D^{ba dc })^*$, which follows from the fact that $W^{\alpha \beta}(z|z')$ is Hermitian.

\section{A classical-quantum Pawula theorem}\label{sec: CQPawulasec}

The inequalities in Equation \eqref{eq: momineq} possess essentially the same structure as the set of inequalities in the classical Pawula theorem \cite{pawula1967rf}, which we review in Appendix \ref{sec: pawulaproof}. However, crucially, they must hold for all $n, m \geq 0$. The difference between the CQ and classical case arises since the zeroth moment of the map $\Lambda^{\mu \nu}(z|z',\delta t)$  is of order $ O(\delta t)$ for the classical-quantum interaction, whilst it is $O(1)$ for the classical case due to the consistency condition at $\delta t=0$. More precisely, for $\delta t=0$ the CQ map in \eqref{eq: CPmap} takes the form $\Lambda^{\mu \nu}(z|z,0) = \delta^{\mu}_0 \delta^{\nu}_0 + O( \delta t)$. As a result, the zeroth moment of the purely classical component of the CQ map is $O(1)$ and so there can be no inequalities relating the zeroth moment of the classical dynamics to any higher order moments, since the zeroth moment always dominates (see Appendix \ref{sec: pawulaproof}). However, for the classical-quantum interaction the zeroth moment is $O(\delta t)$ and so there do exist inequalities relating the zeroth moment to the higher order moments, leading to a strengthened version of the Pawula theorem--which we now state and prove. Recall that non-trivial CQ evolution is one where $W^{\alpha \beta}(z|z')$ is somewhere positive so that the quantum system back-reacts on the classical system,
\newtheorem*{proofss}{CQ Pawula Theorem}\label{proof: Pawula}
\begin{proofss}
For {\it non-trivial CQ evolution}, we must have infinitely many moments defined in Equation \eqref{eq:momentsD}, or else the master equation takes the form 
\begin{align}
\frac{\partial \cqstate(z,t)}{\partial t} & =  \sum_{n=1}^{n=2}(-1)^n\left(\frac{\partial^n }{\partial z_{i_1} \dots \partial z_{i_n} }\right) \left( D^{00}_{n, i_1 \dots i_n} \cqstate(z,t) \right) + \frac{\partial }{\partial z_{i}} \left( D^{0\alpha}_{1, i} \cqstate(z,t) L_{\alpha}^{\dag} \right)  + \frac{\partial }{\partial z_{i}} \left( D^{\alpha 0 }_{1, i} L_{\alpha} \cqstate(z,t) \right)  \nonumber \\ %\label{eq: continuous}
& -i[H(z), \cqstate(z,t)] + D_0^{\alpha \beta}(z) L_{\alpha} \cqstate(z) L_{\beta}^{\dag} - \frac{1}{2} D_0^{\alpha \beta} \{ L_{\beta}^{\dag} L_{\alpha}, \cqstate(z) \}_+  
\end{align}
where $2 D_{2}^{00} \succeq D_1 D_0^{-1} D_1^{\dag}$ and $(\mathbb{I}- D_0 D_0^{-1})D_1 =0$ Here, $D^{-1}_0 $ is the generalized inverse of the matrix $D_0^{\alpha \beta}$, $D_1$ is a matrix in both $\alpha, i$ indices with entries $D_{1,i}^{0 \alpha}$ and $D_2^{00}$ is a matrix in $i,j$ with entries $D_{2,ij}^{00}$. Furthermore, the zeroth moment, $D_0^{\alpha \beta}(z)$ cannot vanish. 
\end{proofss}
 \begin{proof} First, we know from the classical Pawula theorem that the components $D^{00}_{n}$ must vanish for $n \geq 3 $. Now, consider the inequality in \eqref{eq: momineq} for $n,m \geq 1$. Suppose any even CQ moment vanishes, so that $D^{ab cd}_{ 2n} =0$ for all $a,b,c,d$, then so must $D^{ab cd}_{ 2n+m}  =0$, meaning all higher order moments also vanish. Furthermore, if $D^{abcd}_{2n+2m}=0$, for all $a,b,c,d$, then $D^{ab cd}_{ 2n+m}  =0$. Denoting $r=n+m$ we see if $D^{ab cd}_{2r} =0$ then $D^{ab cd}_{ r+n} =0$ for $n= 1 \dots r-1$. To summarise: if any even moment vanishes $D^{ab cd}_{ 2r} =0$ we deduce that all higher order moments $D^{ab cd}_{ 2r + n}$ must vanish, as well as the moments $D^{ab cd}_{ r+n}$ for $n =1 \dots r-1$. Except for the case $r=1$, a moment expansion to order $r+n$ will always contain an even moment and so from repeated application of these properties, if any even moment vanishes then $D^{ab cd}_{ n} =0$ for all $n \geq 3$. This is the usual Pawula theorem, but for the CQ case, we also have the inequality \eqref{eq: momineq} for $n=0, m \geq 1$ which tells us
\begin{equation}\label{eq: pawulah}
(2n)! D^{ab ab}_{ 0} D^{cd cd}_{2m, i_1 i_1 \dots i_{m} i_{m}} \geq |(m)!D^{ab cd}_{ m, i_1 \dots i_{m}} |^2
\end{equation}
We can use this to strengthen the condition. Taking any even moment to be zero we deduce that $D^{ab cd}_{4} =0$. But then from \eqref{eq: pawulah} we must then have $D^{ab cd}_{ 2} =0$, which in turn implies $D^{ab cd}_{ 1}=0$. Hence we see if any even moment vanishes, then all of the moments $D^{\alpha \beta}_n$ $n \geq 1$ vanish. Hence, we conclude that the block $\Lambda^{\alpha \beta}(z|z')$ describes pure quantum evolution.  As a consequence, if any of the even moments greater than two vanish, we can write the transition amplitude of Equation \eqref{eq: block1} in block form as 
\begin{equation} \label{eq: block2}
\Lambda^{\mu \nu}(z|z',\delta t) 
=\resizebox{.8\hsize}{!}{$\begin{bmatrix}
    \delta(z,z') +  \delta t \sum_{n=0}^{2}(-1)^n\left(\frac{\partial^n }{\partial z_{i_1} \dots \partial z_{i_n} }\right) \left( D^{00}_{n, i_1 \dots i_n}(z,\delta t) \delta(z,z') \right)      & \delta t \sum_{n=0}^{1}(-1)^n\left(\frac{\partial^n }{\partial z_{i_1} \dots \partial z_{i_n} }\right) \left( D^{0\alpha}_{n, i_1 \dots i_n}(z,\delta t) \delta(z,z') \right) \\
    \delta t \sum_{n=0}^{1}(-1)^n\left(\frac{\partial^n }{\partial z_{i_1} \dots \partial z_{i_n} }\right) \left( D^{\alpha 0}_{n, i_1 \dots i_n}(z,\delta t) \delta(z,z') \right)  & \delta t D^{\alpha \beta}_0 \delta(z,z')  \\
\end{bmatrix}$},
\end{equation}
where we remember that 
\begin{equation}
    D^{00}_0(z,t) \mathbb{I} + D^{0 \alpha}_0 L_{\alpha} + D^{\alpha 0}_0 L_{\alpha}^{\dag} + D_0^{\alpha \beta}L_{\beta}^{\dag}L_{\alpha} = \mathbb{I}
\end{equation}
from the normalization condition \eqref{eq: prob}. 

We now show \eqref{eq: block2} will be positive, if and only if
\begin{equation}
2 D_2 \succeq D_1D_0^{-1} D_1^{\dag}
\label{eq:diosigen}
\end{equation} 
and 
\begin{equation}\label{eq: schurComplementRequirement}
    (\mathbb{I}- D_0 D_0^{-1})D_1 =0
\end{equation}
 where $D^{-1}_0 $ is the generalized inverse of the matrix with elements $D_0^{\alpha \beta}$, $D_1$ is a matrix in both $\alpha, i$ indices with entries $D_{1,i}^{0 \alpha}$, and $D_2^{00}$ is a matrix in $i,j$ with entries $D_{2,ij}^{00}$.
 
 We start from the the positivity condition of $\Lambda^{\mu \nu}(z|z')$ directly, which states that 
\begin{equation} \label{eq: block5}
\int dz  A^*_{\mu}(z,z')\Lambda^{\mu \nu}(z|z',\delta t) A_{\nu}(z,z') \geq 0 .
\end{equation}
for any $A_{\mu}(z,z')$. This follows from the positivity of $\int dz dz' \Lambda^{\mu}(z|z') P_{\mu}(z,z')$ and using the definition $\Lambda^{\mu \nu}(z|z',\delta t) = U^{\dag}_{\mu \sigma}  \Lambda^{\sigma} U_{\sigma \nu}$. Since the $00$ component of \eqref{eq: block2} contains a delta function $\delta(z,z')$ which is order $O(1)$, whilst all other components are order $O(\delta t)$, Equation \eqref{eq: block5} will be always be positive unless we pick $A_0(z,z') = (z-z')^n b(z,z')$ where $b(z,z)$ is non-zero. Since we know $\Lambda^{00}(z|z')$ and $\Lambda^{\alpha \beta}(z|z')$ are positive, we must consider the case in which $A_{\alpha}$ is non-zero, or else the off-diagonal terms of the block matrix \eqref{eq: block2} do not contribute. The only choice of $A_{\mu}(z,z')$ which gets rid of the leading order $\delta(z,z')$, has well defined distributional derivatives, and keeps the off-diagonal terms is of the form $A_0(z,z') \sim (z-z')b(z,z')$ and  $A_{\alpha} = a_{\alpha}(z,z')$, where $a_{\alpha}(z,z)$ is non-zero.

In the case in which we have many classical degrees of freedom $z_i$, the most general choice which get rid of the leading order $\delta(z,z')$ and keeps the off-diagonal terms is $A_0(z,z') \sim b^i(z,z')(z-z')_i$ for some vector $b_i(z,z')$. For this choice of $A_{\mu}(z,z')$ we find the condition for positivity of $\Lambda^{\mu \nu}(z|z', \delta t)$ is
\begin{equation}\label{eq: inequalitycontMain}
2 b^{i*}(z,z)  D^{00}_{2,ij}b^j(z,z) +  b^{i*}(z,z) D^{0 \alpha}_{1,i} a_{\alpha}(z,z) + a^*_{\alpha}(z,z) D^{\alpha 0}_{1,i} b^i(z,z) + a_{\alpha}^*(z,z)D^{\alpha \beta}_0 a_{\beta}(z,z) \geq 0.
\end{equation}

Defining $D_{2}$ to be the $n\times n$ matrix with elements $D^{00}_{2, ij}$, $D_{1}$ to be the $n \times p $ matrix in $i, \alpha$ with elements $D^{0, \alpha}_{1,i}$ and $D_0$ the $p \times p$ matrix in $\alpha, \beta$ with elements $D^{\alpha \beta}_0$ then \eqref{eq: inequalitycontMain} can be written in the form 
\begin{align} \label{eq: matrixInequalityCont}
[ b^*, \alpha^*]  
\begin{bmatrix}
 2 D_{2} & D_1\\ D^*_1 & D_0 
\end{bmatrix} 
\begin{bmatrix}
 b \\ \alpha
\end{bmatrix} \geq 0 ,
\end{align}
which is equivalent to the condition that the $(n + p)\times (n + p)$ matrix be positive semi-definite
\begin{align} \label{eq: blockMatrixPositivity}
    \begin{bmatrix}
 2 D_{2} & D_1\\ D^*_1 & D_0 
\end{bmatrix} 
\succeq 0.
\end{align}
Since $D_0$ and $D_2$ must be positive semi-definite and since \eqref{eq: blockMatrixPositivity} is a block matrix we know that the Schur complement of $D_0$ must be positive semi-definite, which implies Equations \eqref{eq:diosigen}, \eqref{eq: schurComplementRequirement}.
From Equation \eqref{eq: schurComplementRequirement} we see that if $D_0$ vanishes, then so does $D_1$ and so we must have decoherence for non-trivial CQ evolution. The master equation then takes the form in Equation \eqref{eq: continuous}. 
In the case of a single Lindblad operator both the trade-off of Equation \eqref{eq:diosigen} and Equation \eqref{eq: continuous} reduce to that found in \cite{diosi1995quantum}.
 We further show in Appendix \ref{sec: arblind} that the Lindblad operators in \eqref{eq: continuous} can be arbitrary, rather than requiring them to be traceless and orthogonal, and the map will still be completely positive -- so long as the positivity conditions on the moments, Equation \eqref{eq:diosigen}, are satisfied.
 
\end{proof} 
This gives us a strengthened version of the Pawula theorem for the CQ couplings -- we either have infinitely many terms in the moment expansion or else the dynamics must take the form of \eqref{eq: continuous}, which is the most general form of master equation with almost surely (a.s) continuous classical trajectories.
Indeed, for classical Markovian dynamics (if we assume the short time moment expansion exists) the only time continuous Markovian process, in the sense that
\begin{equation}\lim _{t \downarrow s} \frac{1}{t-s} \int_{|z-z'|>\delta} \mathrm{d}z \ p(t, z | s, z') =0, \ \ \ \forall \delta > 0,  \end{equation} 
is given by a diffusion process with dynamics described by the Fokker-Plank equation~\cite{risken1989fpe}. In \cite{UCLContinousUnraveling} we show explicitly how the continuous master equation can be unraveled \cite{Belavkin, Dalibard, Gardiner} in terms of stochastic differential equations with (a.s) continuous classical trajectories.

\section{Discussion}\label{sec: conclusion}
In this work, we have introduced the most general Markovian form of a continuous classical-quantum master equation, given by Equation \eqref{eq: continuous}. Any other master equation necessarily causes discrete, finite sized jumps in phase space, or else it violates conservation of probabilities, or fails to be completely positive on the quantum system. Complete positivity is required to ensure the probabilities of measurement outcomes remain positive throughout the dynamics. We achieved this by introducing a classical-quantum Cauchy-Schwarz inequality, which enabled us to derive various inequalities relating the moments of the transition amplitude which appear in the master equation. This allowed us to derive an extended version of the Pawula theorem for CQ dynamics: the Kramers-Moyal expansion must contain an infinite number of terms, or else must be of the form in \eqref{eq: continuous}. We hope that this provides a useful reference for the study of hybrid classical-quantum dynamics in the future. Indeed, in the context of classical-quantum theories of gravity \cite{diosi2011gravity,poulinKITP2,oppenheim_post-quantum_2018, Oppenheim:2020ogy,Oppenheim:2020wdj}, if the space-time metric undergoes continuous dynamics, then one expects a version of equation \eqref{eq: continuous} to generate it. To this end, the field theoretic version of Equation \eqref{eq: continuous} is given in \cite{dec_Vs_diff,Oppenheim:2020wdj}.
Constructing consistent theories of CQ general relativity then amounts to an appropriate choice of Lindblad operators, and couplings $D_0, D_1, D_2$, a realisation of which was given in \cite{oppenheim_post-quantum_2018,UCLpathintegral}.
In \cite{UCLpathintegral} we show that one can arrive at a path integral representation for the continuous master equation, which may be useful in understanding whether such dynamics can retain space-time symmetries such as diffeomorphism invariance.

\section*{Acknowledgements}
We would like thank Maite Arcos, Joan Camps, Isaac Layton, Andrea Russo and Andy Svesko for valuable discussions and Lajos Di\'{o}si and Antoine Tilloy for their very helpful comments on an earlier draft of this manuscript. JO is supported by an EPSRC Established Career Fellowship, and a Royal Society Wolfson Merit Award, C.S. and Z.W.D.~acknowledges financial support from EPSRC. This research was supported by the National Science Foundation under Grant No. NSF PHY11-25915 and by the Simons Foundation {\it It from Qubit} Network.  Research at Perimeter Institute is supported in part by the Government of Canada through the Department of Innovation, Science and Economic Development Canada and by the Province of Ontario through the Ministry of Economic Development, Job Creation and Trade.
\bibliography{CQPawulabib}

\bibliographystyle{apsrev}
%\
\appendix

\section{CQ states with continuous classical degrees of freedom}\label{sec: proof}
In order to make sense of the CQ theory when there are continuous classical degrees of freedom, it is useful to attempt to formalize the notion of a CQ state. To that end we shall define a CQ state an an operator valued measure, which formalizes the notion ``to each $z$ we associate a sub-normalized density matrix such that its trace defines a probability distribution over phase space". We then use this to give an argument as to why any completely positive CQ dynamics can be written in the form of \eqref{eq: CPmap}, even in the case where the classical degrees of freedom are continuous \cite{oppenheim_post-quantum_2018}.

Let $\Omega$ be a set and $\mathcal{A}$ be a $ \sigma$ algebra. A map $\cqstate: \mathcal{A} \to S_{\leq 1}( \mathcal{H})$ is called a \textit{CQ state} if 
\begin{enumerate}
\item For each $A \in \mathcal{A}$, $\cqstate(A)$ is a sub-normalised, density operator on $\mathcal{H}$. 

\item  $\cqstate( \emptyset)=0$ and $\cqstate(\Omega)$ is a normalised density matrix. 

\item If $E_{i}$ are disjoint then $\cqstate\left(\bigcup_{j=1}^{\infty} A_{j}\right) =\sum_{j=1}^{\infty} \cqstate\left(A_{j}\right) $

\item  $\mu_{\cqstate}(A) = \Tr{}{ \cqstate(A)}$ defines a probability measure on $\mathcal{A}$
\end{enumerate}

Now from any CQ state $\cqstate$ we can form a real valued measure by setting $\cqstate_{v}(A) = \langle v, \cqstate(A) v\rangle$ for any $v\in \mathcal{H}$. Then, there exists a unique linear map, denoted $f \to \int_{\Omega} f  \cqstate(d \omega)$ with the property that 
\begin{equation}\label{eq: int}
\left\langle v,\left(\int_{\Omega} f \cqstate(d \omega) \right) v \right\rangle=\int_{\Omega} f \cqstate_{v}(d \omega)
\end{equation}
for all bounded measurable complex functions $f: \Omega \to \mathbb{C}$ and all $v \in \mathcal{H}$, where the right hand side of \eqref{eq: int} is the ordinary Lebesgue integral. This follows in the same way as the proof of unique integration for projector valued operators, since a density matrix can be written as a sum of projectors. We can compute the operator valued integral of an arbitrary bounded measurable function as follows. Take a sequence $s_n$ of simple functions converging uniformly to $f$, then the integral of $f$ is the limit, in the operator norm topology, of the integral of the $s_n$.

CQ evolution is a map taking CQ states to CQ states. In order to make sense of this formally, we need to define a notion of measurable super-operators. We first define the space of completely positive super-operators which maps  $S_{\leq 1}( \mathcal{H})$ to itself as $P$. We want to ultimately write
\begin{equation}\label{eq: measureIntegral}
\cqstate'(A) = \int_{\Omega} \Lambda(\omega, A)( \cqstate(d \omega) )
\end{equation}
where $\Lambda: \Omega \times \mathcal{A} \to P$ is such that $\Lambda(\omega, A)$ is a completely positive super-operator for $\omega \in \Omega$ and $A \in \mathcal{A}$. In the rest of the paper, and occasionally in the next subsection, we write the integral in Equation \eqref{eq: measureIntegral} as
\begin{equation}
  \cqstate(z) = \int dz' \Lambda(z|z') \cqstate(z')  , 
\end{equation} and the aim of this section is to give a slightly more precise definition of the integral, so that we can prove a CQ version of Kraus theorem in the case where the classical degrees of freedom are continuous. 

For CQ dynamics, we further ask that $\cqstate'(A)$ defines a CQ state. We give meaning to the integral $\cqstate'(A) = \int_{\Omega} \Lambda(\omega, A)( \cqstate(d \omega) )$ by taking the inner product with a Hilbert space vectors. In particular, if we let $\{|a \rangle\}$ denote an arbitrary basis of vectors in $\mathcal{H}$, and we write the super-operator as $ \Lambda(\omega, A)( \cqstate) = \sum_{abcd} \Lambda^{abcd}(\omega,A) |a \rangle \langle b | \cqstate |c \rangle \langle d| $, then we can make sense of the integral as follows
\begin{equation}\label{eq: inty}
\langle a| \cqstate'(A) |d \rangle  = \sum_{bc} \langle b | \left[ \int_{\Omega} \Lambda^{abcd}(\omega,A) \cqstate(d \omega) \right] |c \rangle
\end{equation}
and we loosely define a measurable CQ dynamics to be a dynamics such that \eqref{eq: inty} is well defined. For simplicity, we assume here that the Hilbert space dimension is finite, but we expect, by analogy with Kraus theorem for quantum operations that this can be extended to any bounded trace-class operation.
\subsection{Proof of Kraus theorem for CQ dynamics}
Here we sketch a proof of a CQ Kraus theorem when the classical degrees of freedom are allowed to be continuous and the Hilbert space is finite dimensional. That is, we given an outline of a proof that every completely positive CQ map can be written in the form of \eqref{eq: CPmap}, with normalization conditions in \eqref{eq: prob}. 

We assume we have a completely positive linear, CQ map $\Lambda$. By linearity, the most general form of the dynamics can be written in the form $\cqstate'(A) =  \int_{\Omega} \Lambda( \omega | A)( \cqstate(d \omega))$ -- we take it given that $\Lambda( \omega | A)$ is measurable in the sense that \eqref{eq: inty} is well defined.

If it is completely positive, then it is certainly $n$ positive. To that end, consider the Choi matrix 
\begin{equation} \label{eq: choi}
\cqstate_{\bar{z}}(A) = \int_{\Omega} \sum_{a b} (I \otimes \Lambda(\omega, A) ) ( E_{ab} \otimes E_{ab} \delta_{\bar{z}}(d \omega)) = \sum_{ab} E_{ab} \otimes \Lambda( \bar{z} | A)( E_{ab}) 
\end{equation}
Where $\delta_{\bar{z}}$ is the delta measure ($\delta_{\bar{z}}(A) =1$ if $\bar{z} \in A$ and $0$ otherwise),  $E_{ab}$ is the natural basis of operators on $\mathcal{H}$, $E_{ab} = |a \rangle \langle b |$ and $\bar{z} \in \Omega$. Since $\sum_{ab} E_{ab} \otimes E_{ab} \delta_{\bar{z}} $ is positive, and $\Lambda$ is assumed to be a completely positive CQ evolution map, $\cqstate_{\bar{z}}(A)$ defines a CQ state on $\mathcal{H}_R \otimes \mathcal{H}$, where $\mathcal{H}_R$ is a reference Hilbert space. Hence, for each $A \in \mathcal{A}$, $\cqstate_{\bar{z}}(A)$ can be diagonalized
\begin{equation}
\cqstate_{\bar{z}}(A) = \sum_{\mu} \lambda^{\mu}( \bar{z}, A) |\phi_{\mu}( \bar{z}, A) \rangle \langle \phi_{\mu}( \bar{z}, A) |
\end{equation}
where the eigenvalues $\lambda^{\mu}( \bar{z}, A)$ are positive for each $A, \bar{z}$. Here $ |\phi_{\mu}( \bar{z}, A) \rangle $ is an element of the product Hilbert space $\mathcal{H}_R \otimes \mathcal{H}$. Now we can find the map $\Lambda( \bar{z}|A)( E_{ab})$ by projection of the Choi matrix on the reference system
\begin{equation}
\Tr{R}{(E_{ba} \otimes I)  \cqstate_{\bar{z}}(A)} = \langle a_R| \cqstate_{\bar{z}}(A) |b_R \rangle = \Lambda( \bar{z} | A)( E_{ab}) 
\end{equation}
If we define the operator $V_{\mu}: \mathcal{H} \to \mathcal{H}$ via its action on the basis of $\mathcal{H}$, $\{ |a \rangle\}$, via $V_{\mu}( \bar{z}, A ) |a \rangle = \langle a_R | \phi_{\mu} ( \bar{z}, A ) \rangle $ then 
\begin{equation}
\Lambda( \bar{z} | A)( E_{ab}) = \sum_{\mu} \lambda^{\mu} ( \bar{z}, A) V_{\mu}( \bar{z}, A ) E_{ab} V_{\mu}^{\dag}( \bar{z}, A )
\end{equation}
Since $\bar{z}$ is arbitrary and $E_{ab}$ is a basis of operators on $\mathcal{H}$ we conclude 
\begin{equation}
\Lambda( \omega | A) = \sum_{\mu} \lambda^{\mu} ( \omega, A) V_{\mu}( \omega, A ) \odot V_{\mu}^{\dag}( \omega, A )
\end{equation}
for all $\omega \in \Omega$. Hence, we can write any complete positive, measurable, CQ dynamics in the form 
\begin{equation} \label{eq: linear}
\cqstate'(A) =  \int_{\Omega} \Lambda( \omega | A)( \cqstate(d \omega)) = \sum_{\mu} \int_{\Omega} \lambda^{\mu} ( \omega , A) V_{\mu}( \omega, A ) \cqstate(d \omega) V_{\mu}^{\dag}( \omega , A )
\end{equation}
and so the CQ map can be written in  the form of \eqref{eq: CPmap}. 

We can recover the normalization conditions in \eqref{eq: prob} as follows. We first note, since the $|\phi_{\mu}( \bar{z}, A) \rangle$ are orthogonal, so are the matrices $V_{\mu}( \bar{z}, A) $ 
\begin{equation}
\sum_{a,b} \langle \phi_{\mu}( \bar{z}, A) | a_R, b \rangle \langle a_R ,b |\phi_{\nu}( \bar{z}, A) \rangle = \Tr{\mathcal{H}}{V_{\mu}^{\dag}( \bar{z}, A) V_{\nu}( \bar{z}, A) }= \delta_{\mu \nu}
\end{equation}
Finally, we note that 
\begin{equation}
\sum_{\mu} \lambda^{\mu}(\bar{z}, A) V_{\mu}^{\dag}( \bar{z}, A) V_{\mu}(\bar{z},A) = \sum_{\mu a b} \lambda^{\mu} ( \bar{z},A) E_{ab}   \langle \phi_{\mu}(  \bar{z}, A) | (E_{ab} \otimes I)|\phi_{\mu}(  \bar{z}, A)  \rangle = \sum_{ab} \Tr{\mathcal{H}_R \otimes \mathcal{H}}{ (E_{ab} \otimes I )\cqstate_{\bar{z}}(A) } E_{ab}
\end{equation}
which using equation \eqref{eq: choi} gives
\begin{equation}
\sum_{\mu} \lambda^{\mu}(\bar{z}, A) V_{\mu}^{\dag}( \bar{z}, A) V_{\mu}(\bar{z},A) = \sum_{\mu} \lambda^{\mu}( \bar{z},A) I
\end{equation}
Since, $\Tr{\mathcal{\mathcal{H}_R \otimes \mathcal{H}}}{ \cqstate_{\bar{z}}(A) } = \sum_{\mu} \lambda^{\mu}( \bar{z}, A) $ defines a probability measure on $\mathcal{A}$ we deduce that 
\begin{equation}
\int_{\Omega} \sum_{\mu} \lambda^{\mu}(\bar{z}, d \omega)  V_{\mu}^{\dag}( \bar{z}, \omega) V_{\mu}(\bar{z},\omega) = I
\end{equation}
which is the normalization condition in \eqref{eq: prob}.

The main lesson here, is that, once we treat the CQ state as an operator valued measure, then the intuition of $\Lambda(z|z')$ as describing a quantum operator for each z,z', holds true, even when the degrees of freedom are continuous.

\section{Classical Pawula theorem}\label{sec: pawulaproof}
In this section, we prove that the classical Pawula theorem follows from the fact $\Lambda^{00}(z|z',\delta t) = \delta(z,z') + \delta t W^{00}(z|z')$ must be positive. 
\newtheorem*{testy}{Pawula Theorem}
\begin{testy}
The series of moments $D^{00}_n$, $n \geq 1$, appearing in the Kramers-Moyal expansion of \eqref{eq: expansion} either contains infinitely many terms, or it truncates after second order, in which case we have a Fokker-Plank equation.
\end{testy}
\begin{proof}
We start from the generalized Cauchy-Schwarz inequality
\begin{equation}
\left[\int f(\Delta) g(\Delta) P(\Delta) \mathrm{d} \Delta\right]^{2} \leq \int f^{2}(\Delta) P(\Delta) \mathrm{d} \Delta \int g^{2}(\Delta) P(\Delta) \mathrm{d} \Delta
\end{equation}
which holds for any non-negative distribution $P(\Delta)$ and arbitrary real valued functions $f(\Delta), g(\Delta)$. Using this with
\begin{equation}
P(\Delta) = \Lambda^{00}(z + \Delta|z, \delta t), \quad f(\Delta) = \Delta_{i_1} \dots \Delta_{i_n} , \quad g(\Delta) = \Delta_{i_{n+m}} \dots \Delta_{i_{2n+2m} }
\end{equation}
gives the inequalities
\begin{equation}
(M^{00}_{2n+m , i_{1}\dots i_{2n+m}})^2 \leq M_{2n, i_{1}i_1 \dots i_n i_n }^{00} M_{2n+2m, i_{n+m} i_{n+m} \dots i_{2n + 2m} i_{2n+2m} }^{00}
\label{eq: moment}
\end{equation}
where $M^{00}_{n, i_1 \dots i_n}(z,\delta t)$ is defined in equation \eqref{eq: moments}.
To prove the Pawula theorem we first relate the coefficients $M^{00}_{n,i_1 \dots i_n}(z,\delta t)$ to the short time expansion coefficients which appear in the master equation. Recall, we have  $M^{00}_{n, i_{1} \dots i_{n} }(z,\delta t) = \delta_n^0 + n! D^{00}_{n_{i_1, \dots i_n } }(z) \delta t + O(\delta t^2)$. We have to be a little careful since $M^{00}_n(z,\delta t) = O(\delta t)$ for $n \geq 1$ but $O(1)$ for $n=0$ \footnote{We use the simplifying notation $M^{00}_n(z,\delta t)$, which means the matrix with components $M^{00}_{n, i_1 \dots i_n}(z,\delta t)$ }. For $n=m=0$ the inequality in \eqref{eq: moment} is trivially satisfied, whilst for $n=0, m\geq 1$ we have no constraints on the short time expansion coefficients since the right hand side of equation \eqref{eq: moment} is $O(\delta t)$ whilst the left hand side is $O(\delta t^2)$. For $n \geq 1, m \geq 0$ we find
\begin{equation}\left[(2 n+m) ! D^{00}_{2n+m , i_{1}\dots i_{2n+m}}\right]^{2} \leq(2 n) !(2 n+2 m) ! D^{00}_{2n, i_{1}i_1 \dots i_n i_n }D^{00}_{2n+2m, i_{n+m} i_{n+m} \dots i_{2n + 2m} i_{2n+2m} 
}\label{eq: paw}
\end{equation}
This gives the usual Pawula theorem, which tells us if any even moment vanishes then all moments with $n \geq 3$ must also vanish. To see this we observe if any even moment vanishes, so that $D^{00}_{2n}=0$, then $D^{00}_{2n +m} =0$ for all $m$. Hence, if any even moment is zero, all of the higher order moments must also vanish. Furthermore, if $D^{00}_{2n+2m} =0$ then it can be seen from \eqref{eq: paw} that $D^{00}_{2n+m}=0$. Denoting $r= n+m$, then this says $D^{00}_{2r}=0$ implies $D^0_{r+n}=0$ for $n=1 \dots r-1$. Hence if any even moment vanishes, $D^0_{2r} =0$, we deduce all higher order moments $D^{00}_{2r+n}$ must vanish, as well as the moments $D^{00}_{r+n}$ for $n=1 \dots r-1$. Except for the case $r=1$, $r+n$ will always contain an even number and so from repeated application of this property we deduce $D^{00}_n$ must vanish for $n \geq 3$.
\end{proof}

\section{Continuous CP evolution with arbitrary Lindblad operators}\label{sec: arblind}
We have shown that any continuous CP CQ map can be written in the form \eqref{eq: continuous} where the Lindblad operators are traceless. We now show that one can pick arbitrary Lindblad operators, $L_{\alpha}$, in \eqref{eq: continuous} and the map will still be completely positive. In other words, we show that Equation \eqref{eq: continuous} where the Lindblad operators are arbitrary is also completely positive, so long as the moments satisfy the positivity conditions.

We first write the arbitrary Lindblad operators, $L_{\alpha}$, in terms of a set of traceless matrices $L_{\alpha} = \bar{L}_{\alpha} + b_{\alpha} \mathbb{I}$. The equation then takes the same form 

\begin{align}\nonumber
\frac{\partial \cqstate(z,t)}{\partial t} & =  \sum_{n=1}^{n=2}(-1)^n\left(\frac{\partial^n }{\partial z_{i_1} \dots \partial z_{i_n} }\right) \left( D^{00}_{n, i_1 \dots i_n} \cqstate(z,t) \right) + \frac{\partial }{\partial z_{i}} \left( D^{0\alpha}_{1, i} \cqstate(z,t) \bar{L}_{\alpha}^{\dag} \right)  + \frac{\partial }{\partial z_{i}} \left( D^{\alpha 0 }_{1, i} \bar{L}_{\alpha} \cqstate(z,t) \right)   \\ 
& -i[H(z), \cqstate(z,t)] + + D_0^{\alpha \beta}(z) \bar{L}_{\alpha} \cqstate(z) \bar{L}_{\beta}^{\dag} - \frac{1}{2} D_0^{\alpha \beta} \{ \bar{L}_{\beta}^{\dag} \bar{L}_{\alpha}, \cqstate(z) \}_+  
\end{align}    
but with a re-scaled Hamiltonian 
\begin{equation}
H(z) \to H(z) + \frac{1}{2i } ( D^{\alpha \beta}_0 b^*_{\beta} \bar{L}_{\alpha} - D^{\alpha \beta}_0 b_{\alpha} \bar{L}^{\dag}_{\beta} ) 
\end{equation}
and a re-scaled \textit{classical} drift coefficient 
\begin{equation}
D_{1,i}^{00} \to D^{00}_{1,i} +  D^{0 \alpha}_{1, i} b^*_{\alpha} + D^{\alpha 0}_{1,i} b_{\alpha}
\end{equation}
We can then write the traceless Lindblad operators in terms of a basis of traceless Lindblad operators $\bar{L}_{\alpha} = V_{\alpha}^{\beta} \tilde{L}_{\beta}$, where $V_{\alpha}^{\beta}$ is inevitable since $\tilde{L}_{\beta}$ form a basis for the traceless operators. Defining $\tilde{D}^{00}_{n,i_1 \dots i_n} =D^{00}_{n,i_1 \dots i_n} $, $\tilde{D}^{\beta 0}_{1,i} = D^{\alpha 0}_{1,i} V^{\beta}_{\alpha}$ and $\tilde{D}^{\gamma \sigma}_0 = V^{\alpha}_{\gamma} D^{\gamma \sigma}_0 (V^{\dag})^{\beta}_{\sigma} $ we find the master equation takes the form
\begin{align}\nonumber
\frac{\partial \cqstate(z,t)}{\partial t} & =  \sum_{n=1}^{n=2}(-1)^n\left(\frac{\partial^n }{\partial z_{i_1} \dots \partial z_{i_n} }\right) \left( \tilde{D}^{00}_{n, i_1 \dots i_n} \cqstate(z,t) \right) + \frac{\partial }{\partial z_{i}} \left( \tilde{D}^{0\alpha}_{1, i} \cqstate(z,t) \tilde{L}_{\alpha}^{\dag} \right)  + \frac{\partial }{\partial z_{i}} \left( \tilde{D}^{\alpha 0 }_{1, i} \tilde{L}_{\alpha} \cqstate(z,t) \right)   \\ 
& -i[H(z), \cqstate(z,t)] + + \tilde{D}_0^{\alpha \beta}(z) \tilde{L}_{\alpha} \cqstate(z) \tilde{L}_{\beta}^{\dag} - \frac{1}{2} \tilde{D}_0^{\alpha \beta} \{ \tilde{L}_{\beta}^{\dag} \tilde{L}_{\alpha}, \cqstate(z) \}_+
\end{align}    
which is now of the form in \eqref{eq: continuous}. Furthermore,
\begin{equation}
    2 \tilde{D}_{2} = 2D_{2} \succeq D_1 \tilde{D}_0^{-1} \tilde{D}_1^{\dag} = D_1 D_0^{-1} D_1^{\dag}
\end{equation}
where we have used the invertibility of $V^{\alpha}_{\beta}$. Hence any equation of the form \eqref{eq: continuous} with arbitrary Lindblad operators and coefficients satisfying the positivity conditions will be completely positive.

\section{A classical oscillator coupled to a quantum one}\label{app: example}

A simple example of the continuous master equation of Equation \eqref{eq: continuous} is given by a classical oscillator coupled to a quantum one. The classical oscillator we describe by the classical Hamiltonian \begin{equation}
    H_c=\frac{1}{2}{p^2}+\frac{1}{2}\omega_c^2q^2,
\end{equation}
and the quantum oscillator we describe by the quantum Hamiltonian 
\begin{equation}
    H_q=\frac{1}{2}{P^2}+\frac{1}{2}\omega_q^2Q^2.
\end{equation}
We take the coupling to be via the interaction Hamiltonian $H_{cq}=D_1qQ$. Then the deterministic part of the dynamics is given by
\begin{align}
\frac{\partial \rho}{\partial t}&=
\{H_c,\cqstate\}
-i[H_q,\cqstate]
-iD_1q[Q,\cqstate]+\frac{1}{2}D_1\{qQ,\cqstate\}-\frac{1}{2}D_1\{\cqstate,qQ\}
\nonumber\\
&=
\{H_c,\cqstate\}
-i[H_q,\cqstate]
-iD_1q[Q,\cqstate]+\frac{1}{2}D_1\left(Q\frac{\partial \cqstate}{\partial p}
+\frac{\partial \cqstate}{\partial p}Q\right).
\label{eq:deterministic_2HO}
%+D_2\frac{\partial^2\rho}{\partial p^2}+\frac{\gamma}{m}\frac{\partial (p \rho)}{\partial p}
\end{align}
The $D_1$ term is the back-reaction of the quantum oscillator on the classical one, and is what we are interested in. 
However, Equation \eqref{eq:deterministic_2HO} is not completely positive, without adding decoherence and diffusion. I.e. we require
\begin{align}\label{eq: Meexample}
\frac{\partial \rho}{\partial t}=&
\{H_c,\cqstate\}
-i[H_q,\cqstate]
-iD_1q[Q,\cqstate]+\frac{1}{2}D_1\left(Q\frac{\partial \cqstate}{\partial p}
+\frac{\partial \cqstate}{\partial p}Q\right)
\nonumber\\&+\lambda\frac{1}{2}[Q,[\cqstate,Q]]
+
D_2\frac{\partial^2\rho}{\partial p^2}
+\gamma\frac{\partial (p \rho)}{\partial p},
\end{align}
where complete positivity requires the decoherence-diffusion trade-off 
\begin{align}
D_2\geq \frac{D_1^2}{\lambda}.
\end{align}
 Roughly speaking, during the time that the system is coherent, the diffusion has to mask the force that the quantum system exerts on the classical one.  
 The master Equation \eqref{eq: Meexample} is of the form originally studied by Diosi \cite{diosi1995quantum} with the addition of a friction term with coupling $\gamma$ which enables one to dampen the effect of the diffusion.

\end{document}